\documentclass[final,,twoside]{IEEEtranTCOM}
\normalsize
\ifCLASSINFOpdf
\else
\fi

\usepackage{amsthm}
\usepackage{amsmath, bm}
\usepackage{amssymb}
\usepackage{graphicx}
\usepackage{esint}
\usepackage{amsfonts}
\usepackage{cite}
\usepackage{balance}
\usepackage{caption}
\usepackage{subcaption}
\usepackage{epstopdf}
\usepackage{color}
\usepackage{lettrine}
\makeatletter

\theoremstyle{plain}
\newtheorem{thm}{\protect\theoremname}

\makeatother

\providecommand{\theoremname}{Theorem}

\theoremstyle{plain}

\makeatother

\providecommand{\lemmaname}{Remark}

\begin{document}
\title{Physical Layer Security in the Presence of Interference}

\author{Dimitrios~S.~Karas,~\IEEEmembership{Member,~IEEE}, Alexandros-Apostolos~A.~Boulogeorgos, \IEEEmembership{Member,~IEEE},  George~K.~Karagiannidis,~\IEEEmembership{Fellow,~IEEE}, and Arumugam Nallanathan,~\IEEEmembership{Fellow,~IEEE}

\thanks{D. S. Karas, A.-A. A. Boulogeorgos, and G. K. Karagiannidis are with the Department of Electrical and Computer Engineering, Aristotle University of Thessaloniki, Greece (e-mail: \{dkaras, ampoulog, geokarag\}@auth.gr).}%
\thanks{A. Nallanathan is with the Department of n the Department of Informatics at King's College London, UK (e-mail: arumugam.nallanathan@kcl.ac.uk)}
}


\maketitle	
\begin{abstract}
We evaluate and quantify the joint effect of fading and multiple interferers on the physical-layer (PHY) security of a system consisted of a base-station (BS), a legitimate user, and an eavesdropper. To this end, we present a novel closed-form expression for the secrecy outage probability, which takes into account the fading characteristics of the wireless environment,  the location and the number of interferers, as well as the transmission power of the BS and the interference. 
The results reveal that the impact of interference should be seriously taken into account in the design and deployment of a wireless system with PHY security.
\end{abstract}
\begin{IEEEkeywords}
Interference, 
Secrecy Outage Probability, Physical layer security.
\vspace{-0.1cm}
\end{IEEEkeywords}

\section{Introduction}\label{S:Intro}

\lettrine{P}{ }hysical layer (PHY) security has received significant attention in the last years, since it can provide reliable and secure communication by employing the fundamental characteristics of the transmission medium, such as multi-path fading~\cite{C:Ha2014,J:Karas2016}. As a result, {a} great amount of effort was put in analyzing the performance of such systems. 
Scanning the open literature, most of the related works have neglected the impact of interference and fading on the security performance of wireless systems. However, in modern heterogeneous wireless enviroments, interference {is} an inevitable key factor for the communication system's performance~\cite{A:What_will_5G_be}. 

The above mentioned scenarios motivated a general investigation of the effect of interference on the security performance of wireless systems~\cite{J:Mukherjee2014, J:Yener2015}. Specifically, a scenario where two independent confidential messages are transmitted to their respective receivers (RXs), which interfere with each other was examined in~\cite{J:RLiu2008}. In this work, the equivocation rate at the eavesdropper was used as a metric to ensure mutual information-theoretic secrecy. Furthermore, in~\cite{J:Liang2009}, the problem of security in the presence of interference was examined from a similar point of view, where two transmitters (TXs) sent two messages to a cognitive RX, who should be able to decode both messages, and a non-cognitive RX, which is able to decode only one message, while the other is kept secret. Moreover, in~\cite{C:Shu2011}, a system that consisted of a primary TX-RX pair, as well as a number of secondary transceivers, and a single eavesdropper, was examined. 
However, in~\cite{C:Shu2011}, the impact of multipath fading was neglected. 
In \cite{J:Nguyen2015}, the secrecy capacity was investigated for a cognitive radio system with security based on artificial noise, assuming full channel state information (CSI) knowledge for the legitimate RX's channel, and partial CSI for the eavesdropper's~channel. Finally, the impact of interference on multi-user scheduling transmission schemes was investigated in \cite{C:Zou1} and~\cite{C:Zou2}.
{However, in these works the fading characteristics of the interference channels were not taken into consideration.}

To the best of the authors' knowledge, the joint effect of interference and fading in PHY security has not been {addressed} in the open technical literature. Motivated by this, in this paper, we examine PHY security for a system, where a TX aims to communicate securely with a legitimate RX, in the presence of an eavesdropper. The signals transmitted by an arbitrary number of base-stations (BSs) cause interference in the signals received by the legitimate RX and the eavesdropper. All TXs and RXs are assumed to be equipped with a single antenna. Also, all wireless links are subject to Rayleigh fading, and statistical CSI is assumed for all channels. To this end, a closed-form expression for the secrecy outage probability (SOP) is~derived.

\vspace{-0.3cm}
\section{System and signal model}\label{sec:SSM}
We consider the downlink scenario in a wireless network that consists of a BS, which aims to transmit a confidential message to a legitimate user, in the presence of an eavesdropper, and $M$ other BSs, which operate in the same frequency band, i.e., they are interferers. For convenience, in what follows, we will refer to the BS as Alice ($A$), the legitimate user as Bob ($B$), and the eavesdropper as Eve ($E$). 

The baseband equivalent signals received by $B$ and $E$ can be respectively obtained as
\begin{align}\label{ITS3:BobSignal}
y_B&=h_{B} x + \sum_{i=1}^{M} h_{Bi} x_i+n_B,
\\
y_E&=h_{E} x + \sum_{i=1}^{M} h_{Ei} x_i+n_E,
\label{ITS3:EveSignal}
\end{align}
where $x$ denotes the transmitted signal by $A$, and $x_i$ denotes the transmitted signal by the $i$-th interferer. Also, $n_B$ and $n_E$ are zero-mean complex Gaussian random variables (RVs) that models the  additive white Gaussian noise (AWGN), with power spectral density $N_0$ at both $B$'s and $E$'s RXs.
Moreover, the baseband equivalent channel between $A$ and $B$ is denoted by $h_B$, while the one between $A$ and $E$ by $h_E$. The baseband equivalent channels between the $i$-th interferer and $B$ are denoted by $h_{Bi}$, whereas those between the $i$-th interferer and $E$ by $h_{Ei}$. Due to the distance, $d_{\mathcal{X}}$, between $A$ and node $\mathcal{X}\in\{B, E\}$, the channel gain can be expressed as in~\cite{J:VPoor2014},
$h_{\mathcal{X}}=\frac{g_{\mathcal{X}}}{\sqrt{1+d_{\mathcal{X}}^{\alpha}}},$
where $g_{\mathcal{X}}$ and $\alpha$ denote the  fading channel and the path loss coefficients, respectively. Similarly, the channel gain between the $i$-th interferer and node $\mathcal{X}$ is given by
$h_{\mathcal{X}i}=\frac{g_{\mathcal{X}i}}{\sqrt{1+d_{\mathcal{X}i}^{\alpha}}},$
where $\mathcal{X} \in \{B, E\}$, while $g_{\mathcal{X}i}$ denotes the  fading channel coefficient, and $d_{\mathcal{X}i}$ denotes the distance between the $i$-th interferer and node $\mathcal{X}$. Note that  $g_{\mathcal{X}}$ and $g_{\mathcal{X}i}$ are zero-mean complex Gaussian RVs with variance equals $1$. Hence,  $|g_{\mathcal{X}}|^2$ and $|g_{\mathcal{X}i}|^2$ follow Rayleigh~distribution. 

Based on~\eqref{ITS3:BobSignal} and \eqref{ITS3:EveSignal}, the instantaneous signal to interference and noise ratio (SINR) at $B$ and $E$ can be expressed~as
\begin{equation}\label{ITS3:gamma_BE}
\gamma_{\mathcal{X}} = \frac{ \frac{E_s}{1+d_{\mathcal{X}}^{\alpha}} |g_{\mathcal{X}}|^2 }{N_0+\sum_{i=1}^{M} |g_{\mathcal{X}i}|^2 \frac{E_{si}}{1+d_{\mathcal{X}i}^{\alpha}}},
\end{equation}
where $E_s$ represents the energy of the signal transmitted by Alice, while $E_{si}$ represents the energy of the signal transmitted by the $i$-th interferer.

\vspace{-0.3cm}
\section{Secrecy Outage Probability}\label{ITS2:sec:SOP}
In this section, we evaluate the SOP, which is defined as the probability that the secrecy capacity is lower than a target secrecy rate, $r_s$, i.e.,
$P_o(r_s) = Pr\left( C_{B} - C_{E} \leq r_s \right),$
or
\begin{equation}
P_o(r_s) = Pr\left(\log_{2}\left(\frac{\gamma_B+1}{\gamma_E+1}\right) \leq r_s \right),
\end{equation}
where 
$C_{B} = \log_2\left(\gamma_B+1\right)$
and 
$C_{E} = \log_2\left(\gamma_E+1\right)$
denote the capacity of  A-B and A-E links,~respectively. 

\vspace{-0.2cm}
\begin{thm}\label{ITS3:SOP_theorem}
The SOP can be expressed in closed form as in~\eqref{ITS3:SecOutProb_4}, given at the top of the next page.
\begin{figure*}
\begin{align}\label{ITS3:SecOutProb_4}
&P_o(r_s) = 1  - \frac{E_{s} N_0}{2^{r_s} (1+d_B^{\alpha})} e^{\left(\frac{1}{\tilde{\gamma}_B} + \frac{1}{\tilde{\gamma}_E}\right)} 
\nonumber \\ & \times\hspace{-0.1cm}
\sum_{i=1}^{M} \hspace{-0.1cm} 
\sum_{j=1}^{M} \hspace{-0.1cm}
\frac{\Xi_B(i) \Xi_E(j)}{b_{Bi} b_{Ej}} \hspace{-0.1cm}
\left( \hspace{-0.1cm}
\frac{\tilde{\gamma}_E e^{-K} }{(L_{Ej}-1)(L_{Bi}-L_{Ej})} 
\right. \hspace{-0.1cm}
- \hspace{-0.1cm}\frac{K \tilde{\gamma}_E e^{L_{Ej} K} \mathcal{E}i\left(-\left(L_{Ej}+1\right) K\right) }{(L_{Bi}-L_{Ej})}  
\hspace{-0.1cm}
+ 
\hspace{-0.1cm}
\frac{\tilde{\gamma}_E  e^{L_{Bi} K} \mathcal{E}i\left(-\left(L_{Bi}+1\right) K\right) }{L_{Bi}-L_{Ej}}  
\nonumber \\ & \hspace{+1cm} 
 -  \frac{\tilde{\gamma}_E e^{L_{Ej} K} \mathcal{E}i\left(-\left(L_{Ej}+1\right) K\right)}{L_{Bi}-L_{Ej}}  
 +
 \frac{e^{K L_{Bi}}\mathcal{E}i\left( -(1+L_{Bi}) K\right) }{L_{Ej}-L_{Bi}} 
\left.
-  \frac{e^{K L_{Ej}}\mathcal{E}i\left( -(1+L_{Ej}) K\right)}{L_{Ej}-L_{Bi}}  
 \right).
\end{align}
\vspace{-0.6cm}
\hrulefill
\end{figure*}
In~\eqref{ITS3:SecOutProb_4}, 
\begin{align}\label{Eq:K}
K & = \frac{1}{\tilde{\gamma}_B} 2^{-r_s}+\frac{1}{\tilde{\gamma}_E},
\\
L_{Bi} & = \frac{E_s - (1+d_B^{\alpha}) b_{Bi}}{(1+d_B^{\alpha}) 2^{r_s} b_{Bi}},
\label{Eq:L_Bi}
\\
\label{Eq:L_Ej}
L_{Ej} &= \frac{E_s - (1+d_E^{\alpha}) b_{Ej} }{(1+d_E^{\alpha}) b_{Ej}},
\end{align}
while 
$\tilde{\gamma}_B = \frac{E_s}{(1+d_B^a) N_0}$
\text{ and }
$\tilde{\gamma}_E = \frac{E_s}{(1+d_E^a) N_0}.$
Also, $\Xi_{\mathcal{X}}(i)$, $\mathcal{X} \in \{B, E\}$ is defined in \cite[Eqs. (8) and (9)]{J:Karagiannidis2006}\footnote{Note that there is a typo in~\cite[Eqs. (8)]{J:Karagiannidis2006}. The correct expression is provided in~\cite{A:Effects_of_multiple_MUs_in_spectrum_sensing}.} {and $\mathcal{E}i(\cdot)$ is the exponential integral function defined in~\cite[Eq. (5.1.4)]{B:Handbook_of_Mathematical_Functions_with_Formulas_Graphs_and_Mathematical_Tables}. }
\end{thm}

\vspace{-0.1cm}
\begin{IEEEproof}
Please refer to the Appendix.
\end{IEEEproof}

Theorem 1 reveals that the SOP does not only depend on the characteristics of the links between A and B/E, but also on the characteristics of the links between the interferers and B/E, as well as the number of interferers. In other words, Theorem 1 quantifies the importance of taking into account the impact of interference in PHY~security.

\vspace{-0.3cm}
\section{Numerical Results and Discussion}\label{ITS2:sec:Results}

In this section, we evaluate and illustrate the joint effect of fading and interference on the performance of wireless systems with PHY security. 
Unless otherwise stated, we assume that the distance between Alice and Bob is $2.5\text{ }\mathrm{m}$, while the distance between Alice and Eve is $25\text{ }\mathrm{m}$. Also, there are three interfering BSs, and their normalized distances from Bob are $10$, $20$ and $25$, whereas their corresponding normalized distances from Eve are $15$, $10$ and $5$. In all cases, the target secrecy rate $r_s$ is expressed in bit/s/Hz.
Moreover, it is assumed that the signals transmitted by the interferers have equal energy, denoted by~$E_{sI}$.

\begin{figure}
\centering\includegraphics[width=0.67\linewidth,trim=0 0 0 0,clip=false]{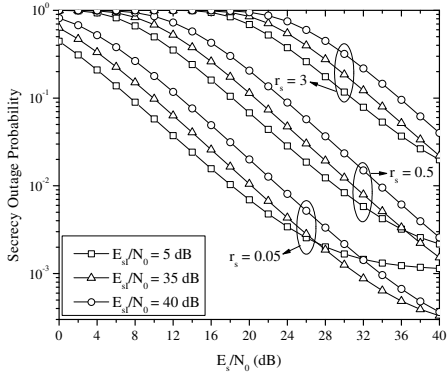}
\vspace{-0.1cm}
\caption{SOP against $E_s/N_0$ for different values of $r_s$ and $E_{sI}/N_0$.}
\vspace{-0.4cm}
\label{fig:ITS2:si_fig1}
\end{figure}
\begin{figure}
\centering\includegraphics[width=0.67\linewidth,trim=0 0 0 0,clip=false]{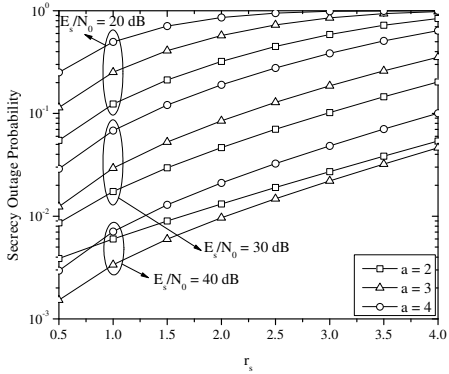}
\vspace{-0.1cm}
\caption{SOP against $r_s$ for different values of $\alpha$ and $E_s/N_0$.}
\vspace{-0.4cm}
\label{fig:ITS2:si_fig2}
\end{figure}
Fig.~\ref{fig:ITS2:si_fig1} depicts the SOP as a function of $E_s/N_0$ for different values of $r_s$ and $E_{sI}/N_0$, and $\alpha = 3$. {We observe that} the SOP decreases as $E_s/N_0$ increases. Furthermore, for given $E_s/N_0$ and $E_{sI}/N_0$, higher rates lead to higher values of the SOP. 
Also, in the examined scenario, in the low $E_s/N_0$ regime, low values for the SOP are achieved if the interferers have low $E_{sI}/N_0$. On the other hand, in the high $E_s/N_0$ regime, low values of the SOP are achieved if the interferers have high~$E_{sI}/N_0$.

In Fig.~\ref{fig:ITS2:si_fig2}, the SOP is illustrated as a function of  $r_s$ for different values of $E_s/N_0$ and  $\alpha$. We observe that, regardless {of} the values of $E_s/N_0$ and $\alpha$, as $r_s$ increases, the SOP also increases. Furthermore, for given $r_s$ and $\alpha$, the increase of $E_s/N_0$ {results in} lower values for the SOP. 
On the other hand, the impact of $\alpha$ on the SOP is not as straightforward. For fixed $E_s/N_0$, $\alpha = 4$ yields the highest SOP in almost all the $r_s$ regime. However, the SOP for $\alpha = 2$ is higher than for $\alpha = 3$ when $E_s/N_0 = 40$ dB, while the SOP for $\alpha = 3$ is higher than for $\alpha = 2$ when $E_s/N_0 = 20$ dB or $E_s/N_0 = 30$ dB. This behavior indicates the dependence of the secrecy performance on the spatial placement of the elements of the system as well as the pathloss parameters.

\begin{figure}
\centering\includegraphics[width=0.67\linewidth,trim=0 0 0 0,clip=false]{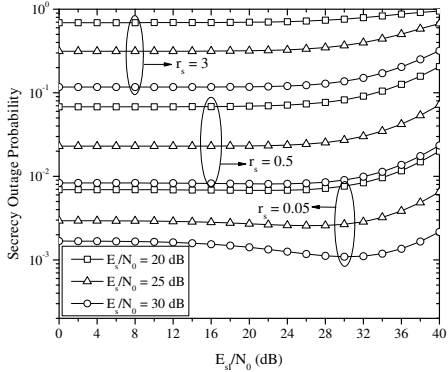}
\vspace{-0.1cm}
\caption{SOP against $E_{sI}$ for different values of $r_s$ and $E_s/N_0$.}
\vspace{-0.4cm}
\label{fig:ITS2:si_fig3}
\end{figure}
\begin{figure}
\centering\includegraphics[width=0.67\linewidth,trim=0 0 0 0,clip=false]{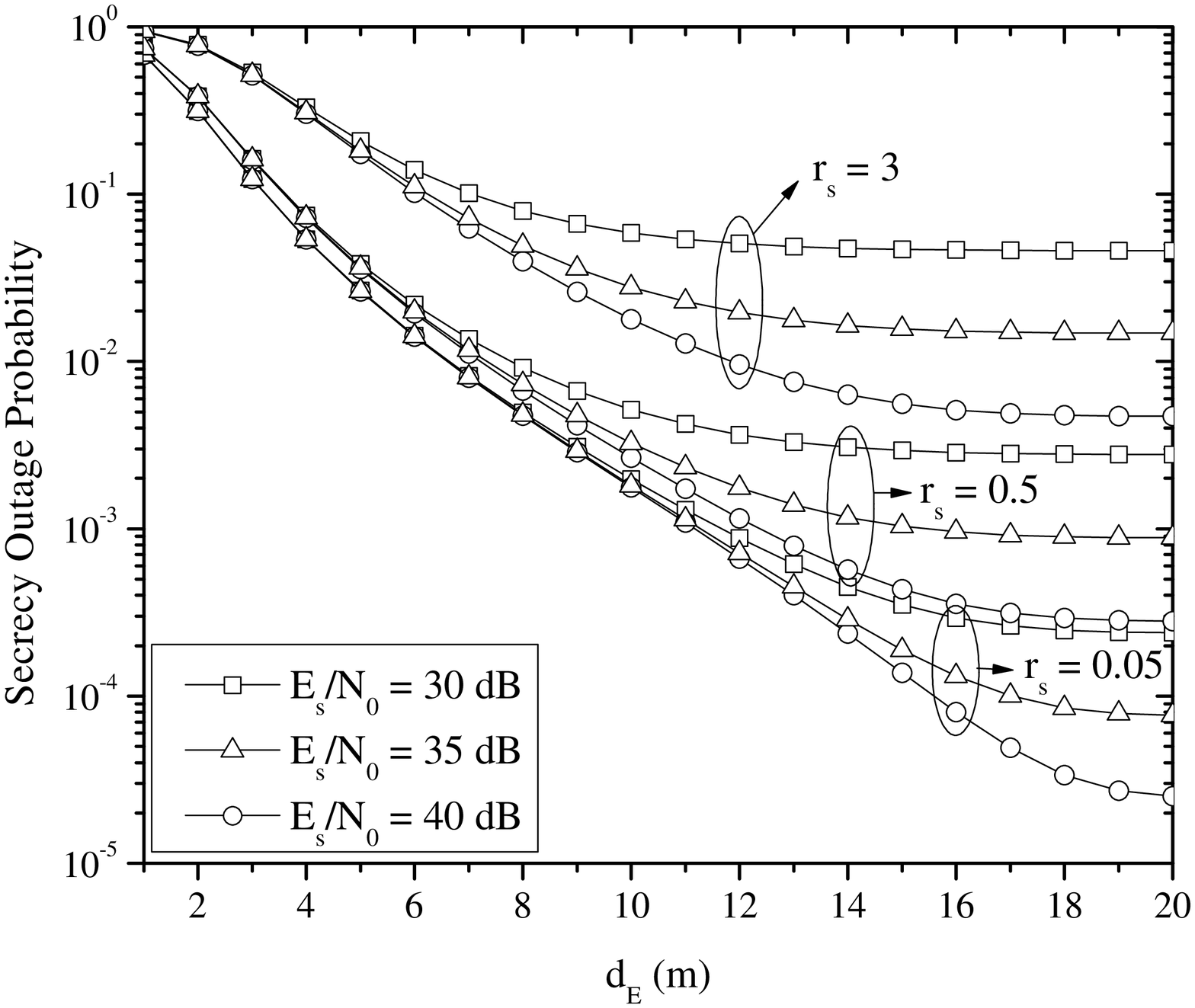}
\vspace{-0.1cm}
\caption{SOP against $d_E$ for different values of $r_s$ and $E_s/N_0$.}
\vspace{-0.4cm}
\label{fig:ITS2:si_fig5}
\end{figure}
\begin{figure}
\centering\includegraphics[width=0.67\linewidth,trim=0 0 0 0,clip=false]{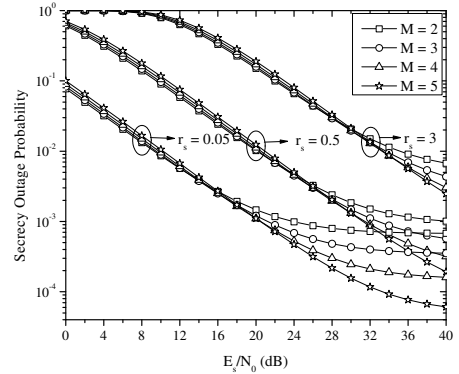}
\vspace{-0.1cm}
\caption{SOP against $E_s/N_0$ for different values of $M$ and $r_s$.}
\vspace{-0.4cm}
\label{fig:ITS2:si_fig6}
\end{figure}
Fig. \ref{fig:ITS2:si_fig3} demonstrates the SOP as a function of $E_{sI}/N_0$ for different values of $r_s$ and $E_s/N_0$, and $\alpha = 3$. Regardless {of} the values of $E_{sI}/N_0$  and $E_s/N_0$, it can be seen that for given $E_{sI}/N_0$  and $E_s/N_0$, as $r_s$ increases, the SOP also increases. However, for given $r_s$, higher values of $E_s/N_0$ lead to a lower SOP. Moreover, it is observed that as $E_{sI}/N_0$ changes, the behavior of the SOP is not straightforward. Specifically, in some cases we observe that as $E_{sI}/N_0$ increases, the SOP decreases until a certain point, and increases afterwards. {This is expected, because the interferers are, on average, closer to Eve than to Bob. Therefore, an increase in $E_{sI}$ is more beneficial to Bob than to Eve. However, as $E_{sI}$ increases, the energy of the signal received by Bob from Alice becomes smaller compared to the energy received from the interferers. Therefore, the capacity of the Alice-Bob and Alice-Eve channels tend to zero, and so does the secrecy capacity, leading to higher values of the SOP.}

Next, we present the impact of interference on PHY security for different positions of Eve. We assume that Alice, Bob and the interferers are placed at fixed locations, while Eve can be placed at $1\text{ }\mathrm{m}$ intervals on a staight line that goes through {Alice and Bob}, up to $20\text{ }\mathrm{m}$ from Alice. Also, in this scenario, $E_{sI}/N_0 = 35$ dB and $\alpha = 3$. In Fig. \ref{fig:ITS2:si_fig5}, we observe that, for a fixed $r_s$, when $d_E$ increases, the SOP decreases. Moreover, we observe that, for fixed $E_s/N_0$ and $d_E$, higher values $r_s$ lead to a higher SOP. In all cases, when Eve moves further from Alice and closer to the interferers, the SOP decreases.

Finally, we investigate the impact of the number of interfering BSs on the SOP. Fig. \ref{fig:ITS2:si_fig6} depicts the SOP as a function of $E_s/N_0$ for different values of $r_s$ and $M$. Also, it was assumed that $E_{sI}/N_0 = 25 dB$ and $\alpha = 3$. The distance between Alice-Bob and Alice-Eve was $d_B = 1$ m and $d_E = 10$ m, respectively. It was assumed that the locations of Alice, Bob, Eve, and the interfering BSs are collinear, and all other elements are on the same side of the line, as defined by Alice's location. The distance of the first interfering BS from Alice was $15\text{ }\mathrm{m}$, and each consecutive BS was placed $1$ m closer to Alice. We observe that, as the value of $E_s/N_0$ increases, the SOP decreases. In the low $E_s/N_0$ regime, a lower number of interferers leads to a lower SOP, but in the high $E_s/N_0$ regime, a larger number of interferers leads to a lower SOP. These results indicate the need to take into consideration the number of interfering BSs in the evaluation of PHY security in a wireless system.

\vspace{-0.5cm}
\appendix
\vspace{-0.2cm}
\section*{Proof of Theorem 1}

The SOP can be expressed as
$P_o(r_s) = P_r\left(\frac{X}{Y} \leq 2^{r_s} \right).$
where 
$X = \gamma_B+1$ 
and
$Y = \gamma_E+1.$
 In order to evaluate the SOP, we first evaluate the cumulative distribution function (CDF) of the SNR at Bob and Eve, which can be obtained~as
\begin{equation}\label{ITS3:cdf_gammaBE_1}
F_{\gamma_{\mathcal{X}}}(x) = \int_{N_0}^{\infty} F_{A}(y x) f_{B}(y) dy,
\end{equation}
where $F_{A_{\mathcal{X}}}(x)$ is the CDF of the RV $A_{\mathcal{X}}$, which is given by 
$A_{\mathcal{X}} = \frac{E_s}{1+d_{\mathcal{X}}^a} |g_{\mathcal{X}}|^2,$
while $f_{B_{\mathcal{X}}}(x)$ is the probability density function (PDF) of the RV $B_{\mathcal{X}}$, which can be expressed as
$B_{\mathcal{X}} = N_0+\sum_{i=1}^{M} |g_{\mathcal{X}i}|^2 \frac{E_{si}}{1+d_{\mathcal{X}i}^{\alpha}}.$
Notice, that $A_{\mathcal{X}}$ and $B_{\mathcal{X}}$ are independent RVs. 

Based on 
\cite{B:Probability_Random_Variables_and_Stochastic_Processes}, $A_{\mathcal{X}}$  follows Rayleigh distribution with CDF given by $F_{A_\mathcal{X}}(x) = 1 - e^{-\frac{1+d_{\mathcal{X}}^{\alpha}}{E_s}x}$. Moreover, since $B_{\mathcal{X}}$ is a weighted sum of Rayleigh distributed RVs, it distribution can be obtained as in~\cite{J:Karagiannidis2006}, and its PDF can be expressed as
\begin{equation}\label{ITS3:den_gammaBE}
f_{B_{\mathcal{X}}}(x) = \sum_{i=1}^{M}\frac{\Xi_{\mathcal{X}}(i)}{b_{\mathcal{X}i}}e^{\frac{x-N_0}{b_{\mathcal{X}i}}},
\end{equation}
where 
$b_{\mathcal{X}i} = \frac{E_{si}}{2(1+d_{\mathcal{X}i}^{\alpha})}.$
The expressions for $b_{Bi}$ and $b_{Ei}$ are used in the definitions of $\Xi_{B}(i)$ and $\Xi_{E}(i)$, respectively. Next, by substituting \eqref{ITS3:den_gammaBE} into \eqref{ITS3:cdf_gammaBE_1}, and after some simplifications, we obtain
\begin{equation}\label{ITS3:cdf_gammaBE_2}
F_{\gamma_{\mathcal{X}}}(x) = 1 - \sum_{i=1}^{M} \frac{\Xi_{\mathcal{X}}(i)}{b_{\mathcal{X}i}} \int_{N_0}^{\infty} \hspace{-0.1cm} e^{\left( -\frac{1+d_{\mathcal{X}}^{\alpha}}{E_s}y x - \frac{y-N_0}{b_{\mathcal{X}i}} \right)} dy.
\end{equation}
By evaluating the integral in \eqref{ITS3:cdf_gammaBE_2}, we obtain
\begin{align}
F_{\gamma_{\mathcal{X}}}\hspace{-0.1cm}(x) \hspace{-0.1cm}= \hspace{-0.1cm}1\hspace{-0.1cm}-\hspace{-0.1cm} 
\sum_{i=1}^{M}\hspace{-0.1cm}\frac{E_{s} \Xi_{\mathcal{X}}(i)e^{-\frac{(1+d_{\mathcal{X}}^{\alpha}) N_0 x}{E_s}}}{E_s + (1+d_{\mathcal{X}}^{\alpha}) b_{\mathcal{X}i} x}.
\end{align}
Next, the CDFs of $X$ can be derived as 
$F_X (x) = F_{\gamma_B}(x-1)$,
or equivalently
\begin{align}\label{ITS3:FX}
F_X(x) = 1- &
\sum_{i=1}^{M}\frac{\Xi_B(i)  E_{s}e^{-\frac{(1+d_B^{\alpha}) N_0 (x-1)}{E_s}}}{E_s + (1+d_B^{\alpha}) b_{Bi} (x-1)}.
\end{align}

Additionally, the PDF of $Y$ can be derived as
$f_Y(x) = \frac{d F_{\gamma_E}(x-1)}{dx}$
which, after some algebraic manipulations, can be rewritten~as
\begin{align}
f_Y(x) &= (1+d_E^{\alpha}) e^{-\frac{(1+d_E^{\alpha}) N_0 (x-1)}{E_s}} 
\nonumber \\&\times 
\sum_{i=1}^{M}\Xi_{E}(i)\left( \frac{E_{s} b_{Ei}}{\left(E_s + b_{Ei} (1+d_E^{\alpha}) (x-1)\right)^2}\right. \nonumber
 \\  & \hspace{1.9cm} \left. + \frac{N_0}{E_s + b_{Ei} (1+d_E^{\alpha}) (x-1)} \right).
 \label{ITS3:fY}
\end{align}

Since $X$ and $Y$ are independent RV, the SOP can be obtained as
\begin{equation}\label{ITS3:SecOutProb_2}
P_o(r_s) = \int_1^{\infty} F_X(2^{r_s}x) f_Y(x) dx.
\end{equation}
By substituting \eqref{ITS3:FX} and \eqref{ITS3:fY} into \eqref{ITS3:SecOutProb_2}, and after some mathematical manipulations, we get
\begin{align}
P_o(r_s) = 1  &-
\sum_{i=1}^{M} \sum_{j=1}^{M} \frac{E_{s} N_0 \Xi_B(i) \Xi_E(j) e^{\left(\frac{1}{\tilde{\gamma}_B} + \frac{1}{\tilde{\gamma}_E}\right)}}{2^{r_s} b_{Bi} b_{Ej}(1+d_B^{\alpha})}   
\nonumber \\ & \hspace{+3cm}\times 
\left( \tilde{\gamma}_E \mathcal{I}_1 +  \mathcal{I}_2  \right),
\label{ITS3:SecOutProb_3}
\end{align}
where $\mathcal{I}_1$ and $\mathcal{I}_2$ can {be} respectively expressed as
\begin{align}
\mathcal{I}_1 =  \int_1^{\infty} \hspace{-0.2cm} \frac{e^{-K y}}{\left(L_{Bi} + y\right) \left(L_{Ej}  + y \right)^2}  dy
\label{Eq:I_1}
\end{align}
\text{ and } 
\begin{align}
\mathcal{I}_2 = \int_1^{\infty}  \frac{e^{-K y}}{\left(L_{Bi} + y\right)\left(L_{Ej} + y \right)} dy.
\label{Eq:I_2}
\end{align}

By setting 
$z=y-1$ 
into~\eqref{Eq:I_1} and~\eqref{Eq:I_2}  and after some basic algebraic manipulations and the use of \cite[Eq.8.359.1]{B:Gra_Ryz_Book}, \eqref{Eq:I_1}  can be  rewritten~as
\begin{align}\label{Eq:I_1_s3}
\mathcal{I}_1 \hspace{-0.01cm}&\hspace{-0.1cm}  =\hspace{-0.1cm}  
\frac{e^{-K} }{(L_{Ej}-1)(L_{Bi}-L_{Ej})} 
\hspace{-0.1cm}-\hspace{-0.1cm} \frac{K \hspace{-0.01cm} e^{ L_{Ej} K}\hspace{-0.01cm} \mathcal{E}i\left(\hspace{-0.05cm}-\hspace{-0.05cm} \left(L_{Ej}+1\right) K\right) }{L_{Bi}-L_{Ej}} 
\nonumber \\ & 
\hspace{-0.5cm}  + \hspace{-0.1cm}   \frac{e^{L_{Bi} K} \mathcal{E}i\hspace{-0.1cm}\left(\hspace{-0.05cm}-\hspace{-0.05cm} \left(L_{Bi}+1\right) K\right) }{L_{Bi}-L_{Ej}}  
\hspace{-0.1cm}    -  \hspace{-0.1cm}   \frac{e^{L_{Ej} K} \mathcal{E}i\hspace{-0.1cm}\left(\hspace{-0.05cm}-\hspace{-0.05cm}\left(L_{Ej}+1\right) K\right)}{L_{Bi}-L_{Ej}}  
\\
\label{Eq:I_2_s3}
\mathcal{I}_2 &\hspace{-0.05cm}  =\hspace{-0.1cm}   \frac{e^{K L_{Bi}} \mathcal{E}i\hspace{-0.1cm}\left( \hspace{-0.05cm}-\hspace{-0.05cm}(1\hspace{-0.05cm}+\hspace{-0.05cm}L_{Bi})\hspace{-0.05cm} K\hspace{-0.05cm}\right) }{L_{Ej}-L_{Bi}} 
\hspace{-0.1cm}- \hspace{-0.1cm}    \frac{e^{K L_{Ej}} \mathcal{E}i\hspace{-0.1cm}\left(\hspace{-0.05cm}-\hspace{-0.05cm}(1\hspace{-0.05cm}+\hspace{-0.05cm}L_{Ej}) \hspace{-0.05cm} K\hspace{-0.05cm}\right)}{L_{Ej}-L_{Bi}}. 
\end{align}

Finally, by substituting ~\eqref{Eq:I_1_s3} and \eqref{Eq:I_2_s3} into~\eqref{ITS3:SecOutProb_3},  we obtain~\eqref{ITS3:SecOutProb_4}.  
This concludes the proof.

\bibliographystyle{IEEEtran}
\bibliography{IEEEabrv,References}
\balance

\end{document}